\documentclass[12pt]{article}
\usepackage[dvips]{graphicx}
\usepackage{pdproc}
\newcommand{\hs}{\ensuremath{\mathit{\Phi}}}
\newcommand{\pbinv}{\ensuremath{\mathrm{pb^{-1}}}}
\newcommand{\fbinv}{\ensuremath{\mathrm{fb^{-1}}}}
\newcommand{\pp}{\ensuremath{p\bar{p}}}
\newcommand{\bb}{\ensuremath{b\bar{b}}}
\newcommand{\nn}{\ensuremath{\nu\bar{\nu}}}
\newcommand{\ttbar}{\ensuremath{t\bar{t}}}
\newcommand{\et}{\ensuremath{E_{\mathrm{t}}}}
\newcommand{\met}{\mbox{\ensuremath{\not \!\! \et}}}
\newcommand{\tb}{\ensuremath{\tan\!\beta}}
\newcommand{\pt}{\ensuremath{p_{\perp}}}
\newcommand{\tautau}{\ensuremath{\tau\tau}}
\newcommand{\cha}{\ensuremath{\widetilde{\chi}}}
\newcommand{\chiz}{\ensuremath{\tilde{\chi}^0}}

  \makeatletter 
  \def\@cite#1{[#1]} 
  \makeatother    
\begin{document}

\renewcommand{\thefootnote}{\alph{footnote}}

\title{
 Searches for Higgs Bosons and Supersymmetry at the Tevatron
}

\author{Volker B\"uscher}

\address{ 
Physikalisches Institut, Universit\"at Freiburg \\
Hermann-Herder-Str. 3, 79104 Freiburg, Germany
%%%%% You may comment out the e-mail address line below.  
%\\ {\rm E-mail: buescher@fnal.gov}}
}

\abstract{
With almost 0.5~\fbinv\ of \pp-collisions delivered in Tevatron
Run~II, both experiments CDF and D\O\ are reporting first results from
a vast number of search analyses.
This article summarizes the current status of
Tevatron searches for Higgs bosons within the Standard Model and its extensions
as well as direct searches for supersymmetric particles. 
Signatures for production of squarks, gluinos, charginos and neutralinos as
predicted by gravity- and gauge-mediated SUSY breaking scenarios have
been considered.
No evidence for Higgs bosons or SUSY particles has been found so far. New limits
are derived which significantly improve on existing limits.
}

\normalsize\baselineskip=15pt

\section{Introduction}
Since March 2001 the Tevatron \pp-collider at the Fermi National
Accelerator Laboratory is operating at a center-of-mass energy
of 1.96~TeV and a bunch spacing of 396~ns (Run~II).
After an initial commissioning period for accelerator and 
detectors, the machine has delivered about 450~\pbinv\ between April
2002 and June 2004. 
Peak luminosities of up to
1.0~$\cdot$~10$^{32}$$\rm{cm}^{-2}\rm{s}^{-1}$ have been achieved so
far.
Once the new recycler ring is fully operational the luminosity is
expected to increase further, resulting in a projected Run~II
dataset of up to 8~\fbinv\ by the year 2009.

Given its high energy and steadily increasing luminosity, the Tevatron
collider is an ideal  
tool for searches for massive particles beyond the reach of LEP.
The results presented in this contribution are based on about 200~\pbinv\ of data 
recorded and analyzed by the two Tevatron experiments CDF and D\O.
After a short description of the two detectors in Section~2, the status of searches for Higgs bosons 
and supersymmetric particles is summarized in Sections~3 and 4. 
For completeness, a brief overview of searches for other new physics is given 
in Section~5.

\section{\label{detectors}CDF and D\O\ Detectors}
The CDF and D\O\ detectors are described in detail in
Ref.~\cite{cdf-detector} and \cite{d0-detector}. 
Only a brief overview is presented in the following.

CDF track reconstruction relies on silicon detectors and a drift chamber
situated inside a solenoid that provides a 1.4~T magnetic field coaxial with
the beam.
The silicon microstrip detector consists of eight cylindrical
layers of mostly double-sided silicon, distributed in radius between
1.5~cm and 28~cm. The system is readout in about 700.000 channels and
can provide 3D precision tracking up to pseudorapidities of 2.0.
Outside of the silicon detectors and for pseudorapidities less than
1.0, charged particles are detected with up to 96 hits per track by
the central outer tracker, an open-cell drift chamber with alternating
axial and 2$^\circ$ stereo superlayers with 12 wires each. 
Just inside the solenoid, a scintillator-based time-of-flight detector
allows particle identification with a timing resolution of about 100~ps.

The electromagnetic (hadronic) calorimeters are lead-scintillator
(iron-scintillator) sampling calorimeters, providing coverage up to
pseudorapidities of 3.6 in a segmented projective tower geometry.
Proportional wire and scintillating strip detectors situated at a
depth corresponding to the electromagnetic shower maximum provide
measurements of the transverse shower profile.
In addition, an early energy sampling is obtained using 
preradiator chambers positioned between the solenoid coil and the
inner face of the central calorimeter.
Outside of the calorimeter and behind additional steel absorbers, a
multi-layer system of drift chambers and 
scintillation counters allows detection of muons for pseudorapidities
up to 1.5.

The tracking system of the D\O\ detector consists of a silicon vertex
detector and 
a scintillating fiber tracker, situated inside a superconducting coil
providing a 2~T magnetic field.
The D\O\ silicon tracker has four cylindrical layers of mostly
double-sided microstrip detectors covering 2.7~cm up to 9.4~cm in
radius, interspersed with twelve disk detectors in the 
central region and four large disks in the forward region. The full system
has about 800.000 channels and provides 3D precision
tracking up to pseudorapidities of 3.0. 
The volume between the silicon tracker and the superconducting coil
is instrumented with eight cylindrical double layers of
scintillating fibers. Each layer has axial and stereo fibers (stereo
angle $\pm 3^{\circ}$) with a diameter of 835~$\mu$m, that are readout
using solid-state photodetectors (Visible Light Photon Counters,
VLPCs).

The D\O\ calorimeter is a Liquid Argon sampling calorimeter with
Uranium absorber (Copper and Steel for the
outer hadronic layers) with hermetic coverage up to 
pseudorapidities of 4.2. 
Signals are readout in cells of projective towers with four
electromagnetic, at 
least four hadronic layers and a transverse segmentation of 0.1 in
both azimuth and pseudorapidity.
The granularity is increased to 0.05 for the third EM layer,
roughly corresponding to the electromagnetic shower maximum.
To provide additional sampling of energy lost in dead material,
scintillator-based detectors are placed in front of the calorimeter
cryostats (preshower detectors) and between the cryostats
(intercryostat detector). The preshower
detectors consist of three layers of scintillator strips with VLPC readout
providing, in addition to the energy measurement, a precise
3-dimensional position measurement for electromagnetic showers.

The D\O\ Muon system consists of three layers of drift tubes and
scintillators, with toroid magnets situated between the first and
second layer to allow for a stand-alone muon momentum measurement. 
Scintillator pixels are used for triggering and rejection of out-of-time
backgrounds in both central and forward region. Proportional drift tubes 
are stacked in three or four decks per layer in the central region. 
Tracking of muons in the forward region is accomplished using decks of
mini drift tubes in each layer, allowing muons to be
reconstructed up to pseudorapidities of 2.0.
The muon system is protected from beam-related backgrounds by
shielding around the beampipe using an iron-polyethylene-lead
absorber.

Both CDF and D\O\ detectors are readout using a three-level trigger system which
reduces the event rate from 2.5~MHz to about 50~Hz. This includes
programmable hardware triggers at Level~1 that provide basic track,
lepton and jet reconstruction, secondary vertex or impact parameter
triggers at Level~2 as well as a PC-based quasi-offline event
reconstruction at Level~3.

\section{\label{higgs}Searches for Higgs Bosons}
As the Run~II luminosity increases, CDF and
D\O\ will start reaching sensitivity to production of low-mass Higgs
bosons beyond the LEP limits. For Standard Model Higgs bosons decaying to \bb, the
production in association with $W$ or $Z$ bosons is the most promising
channel. At masses up to about 180~GeV, Higgs bosons produced via
gluon fusion might be observable in their decays to $WW$.
For models with more than one Higgs doublet, the coupling of the Higgs
boson to b-quarks can be significantly enhanced, allowing to search for
Higgs bosons produced in association with b-quarks.

The following sections summarize the current status and projections
for the most important channels at Tevatron Run~II.

\subsection{\label{s:tev-vh} Associated Production}
The production of Higgs bosons in association with vector bosons can
be searched for in all leptonic decays of $W$ and $Z$: $W\to \ell \nu$,
$Z\to\nn$ and $Z\to \ell\ell$ (with $\ell$=$e$,$\mu$,$\tau$). Sensitivity studies
based on Monte Carlo simulation of detector performance throughout the course
of Run~II exist.\cite{tev-report,tev-hss} 
For $WH$ production, first preliminary
results\cite{tev-smhiggs} of searches in 162~\pbinv\ (CDF) and 174~\pbinv\ (D\O) of 
Run~II data are described in the following.

Final states compatible with the $WH$ signature can be selected by
requiring one isolated lepton and missing transverse energy as well as two
b-tagged jets. 
After additional topological cuts,
backgrounds are entirely
dominated by physics backgrounds from $W/Z$\bb, $WZ$ and \ttbar.
To improve the signal-to-background ratio further, the Higgs boson mass 
has to be reconstructed with the best possible resolution.
Currently, a relative jet energy resolution of 13.9\% has been
achieved by D\O, as measured in Run~II data for central jets at
\et=55~GeV.\cite{tev-hss} It is expected that this can be improved by
30\% due to more sophisticated jet reconstruction algorithms as well
as refinements  
in jet energy calibration, including a calibration of the \bb\ mass
reconstruction using the $Z\to\bb$ signal. 

\begin{figure}[htb]
\begin{center}
\includegraphics*[width=10cm]{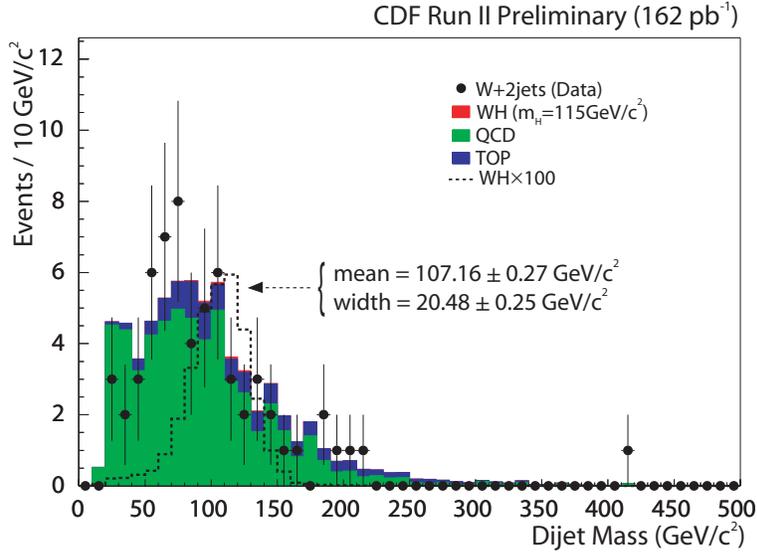}
\caption{%
Invariant \bb\ mass spectrum in search for $WH$ production in
162~$pb^{-1}$ of CDF data in comparison with signal and background expectation.
}
\label{f:tev-wh}
\end{center}
\end{figure}

A Higgs signal is then searched for as an excess in the \bb\ mass
spectrum, as shown in Fig.~\ref{f:tev-wh} 
for the CDF analysis.
No evidence for $WH$ production is observed in current Run~II searches
by CDF and D\O, allowing to set an upper limit on the product of
cross-section and branching fraction
$\sigma$($WH$)$\times$BR($H\to$\bb) of 5~pb for a Higgs boson mass of
120~GeV,\cite{tev-smhiggs} which is still more than an order of magnitude
higher than the Standard Model expectation.

\begin{figure}[htb]
\begin{center}
\includegraphics*[width=10cm]{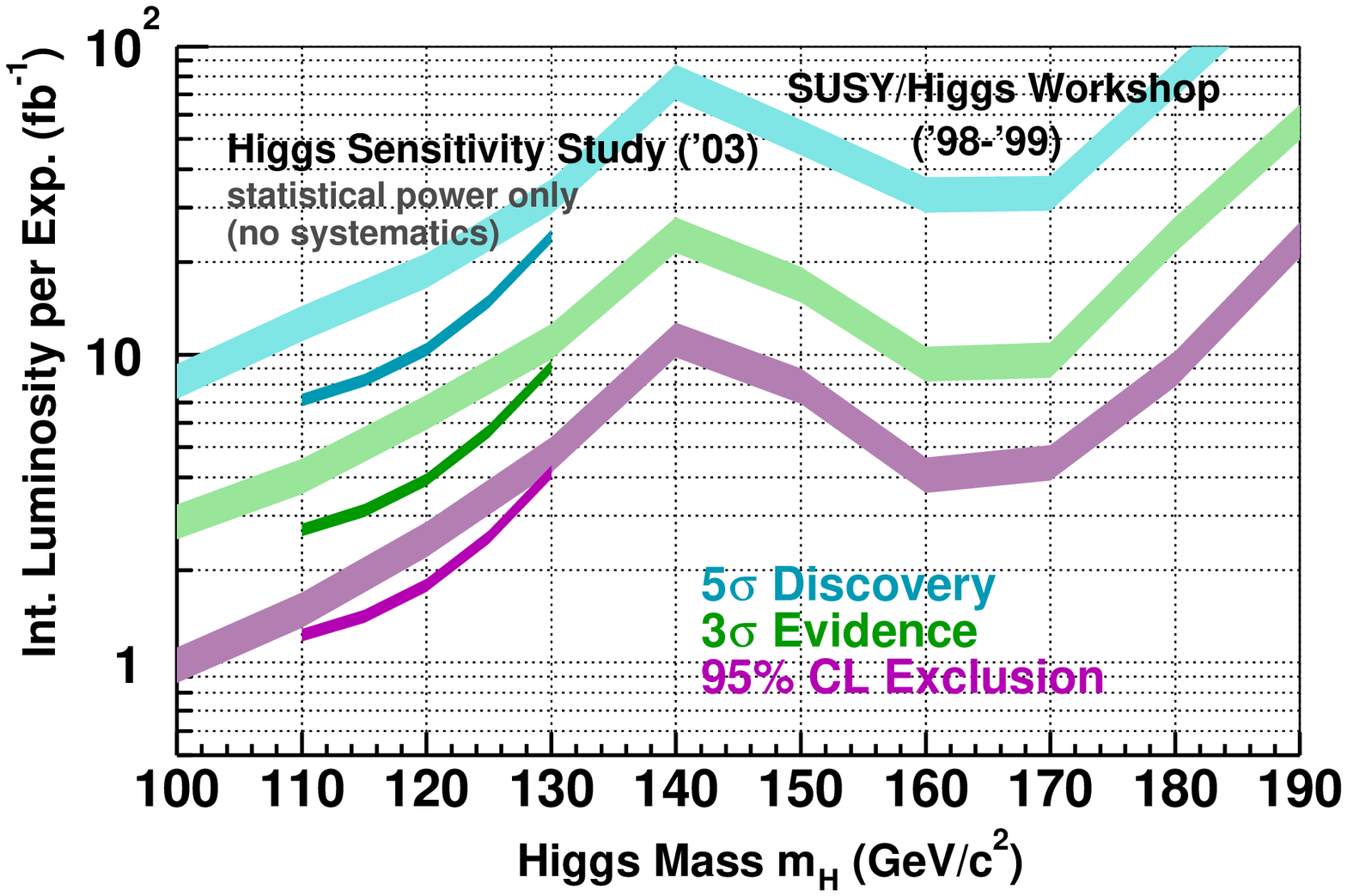}
\caption{%
The integrated luminosity needed per experiment for a 95\% CL
exclusion, a 3$\sigma$ and a 5$\sigma$  
discovery of a Standard Model Higgs boson at the Tevatron as a function of the 
Higgs boson mass, after combining all searches for associated
production with a vector boson. Thin lines show recent estimates based
on tuned full simulation\protect\cite{tev-hss}, thick lines indicate
results of an earlier study with fast
simulation (includes searches for $H\to WW$).\protect\cite{tev-report}.
}
\label{f:tev-higgs}
\end{center}
\end{figure}

In  Fig.~\ref{f:tev-higgs} 
the luminosity required to observe (or exclude) 
a Standard Model Higgs boson is shown as a function of mass.
After combining all channels and both experiments, a sensitivity at
the 95\% C.L. for m$_H$=120~GeV is expected to be achieved with an
integrated luminosity of 1.8~\fbinv\ per experiment. 
Evidence for a signal at 
the 3$\sigma$ (5$\sigma$) level will require 4~\fbinv\ (10~\fbinv) for
the same Higgs boson mass. 
These estimates 
assume a 30\% improvement in jet energy resolution and 
do not include systematic errors.
Given the Higgs event yields (3 $WH$ events selected per \fbinv), these 
analyses will require precise knowledge of the backgrounds over the entire mass
range. While the normalization of the background can be obtained from
a fit outside of the signal region, the shape and relative normalization 
of the \bb\ mass spectrum of the various background components has to be 
known to allow extrapolation below the Higgs peak. Procedures to
obtain this information from data are outlined in
Ref.~\cite{tev-hss} and typically involve a measurement of the
shape of the dijet mass spectrum in background-enriched samples, which
is then extrapolated to the final signal sample using a mixture of
Monte Carlo and data-driven methods.

\subsection{$H\to WW$}
Within the mass range of interest at the Tevatron, Higgs boson decays
into two $W$ bosons are the dominant decay mode for Higgs boson masses
above 140~GeV.  
The relatively clean signature of two leptonic $W$~decays allows to
search for this decay in the gluon-gluon-fusion channel. While the
production cross-section in this channel is higher compared to the
associated production, the suppression due the branching fractions of
the leptonic $W$~decays limits the event yield to only about 4 events per
\fbinv.

\begin{figure}[htb]
\begin{center}
\includegraphics*[width=8cm]{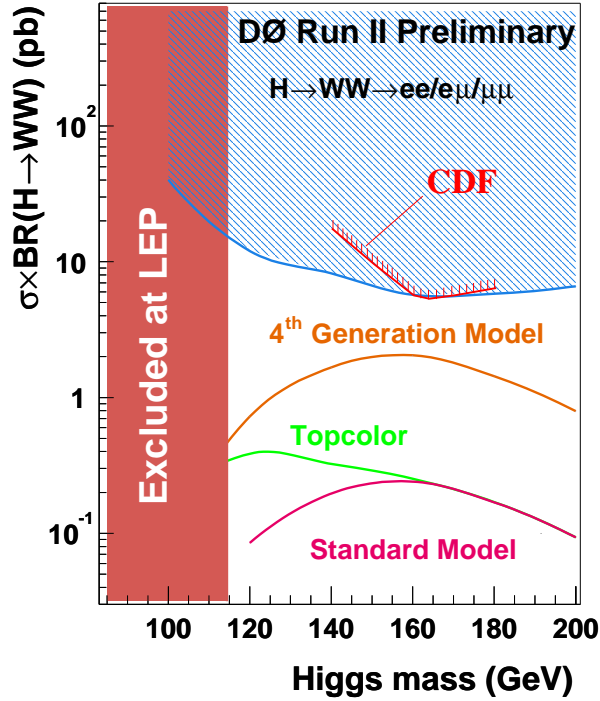}
\caption{%
Upper limit (at 95\% CL) on the cross section $\sigma \cdot BR(H\to
WW)$ set by the D\O\ Collaboration using 176~$pb^{-1}$ of Run~II
data, modified to include the limit set by the CDF Collaboration using 
184~$pb^{-1}$ of Run~II data. The limits are compared to expectations from Standard Model Higgs
production and alternative models.
}
\label{f:tev-hww}
\end{center}
\end{figure}

Both Tevatron collaborations have started analyzing their Run~II data
in search for a $H\to WW$ 
signal.\cite{tev-smhiggs} So far, efficiencies of up to 15--20\% have been achieved for the 
dilepton plus \met\ final states with electrons or muons. The background
is dominated by $WW$ production, which remains after selection cuts with
a cross section of about 25~fb.
Further separation of signal and $WW$ events is
possible using the difference in azimuthal angle~$\Delta\phi$ between the two charged
leptons.\cite{hww-dphi} Due to spin correlations, $\Delta\phi$ tends to be small for
decays of a spin-0 resonance.
Both CDF and D\O\ observe no significant excess of events in 184~\pbinv\ and 
176~\pbinv\ of Run~II data, respectively. Limits on the production
cross-section of $H\to WW$ 
have been set as a function of the Higgs boson mass as shown in
Fig.~\ref{f:tev-hww}. 
For a mass of 160~GeV, cross-sections larger
than 5.6~pb have been excluded at 95\% C.L., which is still more than
an order of magnitude higher than the expectation within the Standard
Model. 

The performance of these analyses is consistent with the
expectations based on the fast simulation, which projected a total
background of 30.4~fb at an efficiency of 18.5\% for a Higgs boson
mass of 150~GeV.\cite{tev-report} Based on this projection,
sensitivity at the 95\% C.L. to a Standard Model Higgs boson with
masses between 160 and 170~GeV will be reached with an integrated
luminosity of 4~\fbinv\ per experiment (10~\fbinv\ for a 3$\sigma$
sensitivity), as shown in Fig~\ref{f:tev-higgs}.

In models beyond the Standard Model, the rate of $H\to WW$ events can be 
enhanced due to larger production cross-sections (models with heavy 4th 
generation quarks) or due to an increase in branching fraction 
(Topcolor models\cite{tev-report}). 
In the former case, the gluon-gluon fusion process is enhanced due to 
loop-diagrams involving heavy quarks by a factor of 
about 8.5 within the mass range of interest at the Tevatron, with only 
a mild dependence on the heavy quark mass.\cite{4gen}
The latter class of models also predicts an enhanced branching fraction 
for $H\to\gamma\gamma$, which can be searched for with diphoton analyses 
to increase the Tevatron sensitivity at low Higgs boson masses.\cite{tev-report}
First Run~II results from D\O\ in this channel exist, but improve only
marginally on existing limits from LEP and Run~I.\cite{tev-diphoton}

\subsection{Neutral Higgs Bosons in Supersymmetry}
Sensitivity to a low-mass Higgs boson is of particular interest within
supersymmetric extensions of the Standard Model, which predict the
existence of at least one neutral Higgs boson~\hs=$h,H,A$ with a mass below
135~GeV. Searches for the Standard Model Higgs boson produced in
association with a vector boson can be
interpreted within SUSY parameter space.
In addition, the enhancement of the Higgs coupling to \bb\ at large \tb\ 
results in sizeable cross-sections for two search channels that are
inaccessible within the Standard Model: the production of Higgs bosons in
association with one or more b-quarks\cite{bh-bbh} as well as the 
gluon-gluon-fusion channel $gg\to \hs$ with the subsequent 
decay $\hs\to\tau\tau$. 

\subsubsection{$\hs b(b)\to bbb(b)$}
The D\O\ collaboration has analyzed 131~\pbinv\ of Run~II data
collected with multijet triggers optimized for the $\bb\hs\to\bb\bb$
signal.\cite{tev-susyhiggs} 
Requiring two jets with transverse momenta~\pt$>$25~GeV and a third
jet with \pt$>$15~GeV, this trigger consumed less than 4~Hz of
Level-3 bandwidth at instantaneous luminosities of \mbox{4.0 $\cdot$
10$^{31}$ $\rm{cm}^{-2}\rm{s}^{-1}$}, while 
maintaining a signal efficiency of about 70\% after offline cuts.
The offline analysis
requires at least three b-tagged jets with \pt$>$15~GeV. Depending on
the Higgs mass hypothesis, the \pt\ cuts for the leading two jets are
tightened to values between 35 and 60~GeV to optimize for best
expected sensitivity.  

\begin{figure}[htb]
\begin{center}
\includegraphics*[width=0.48\textwidth]{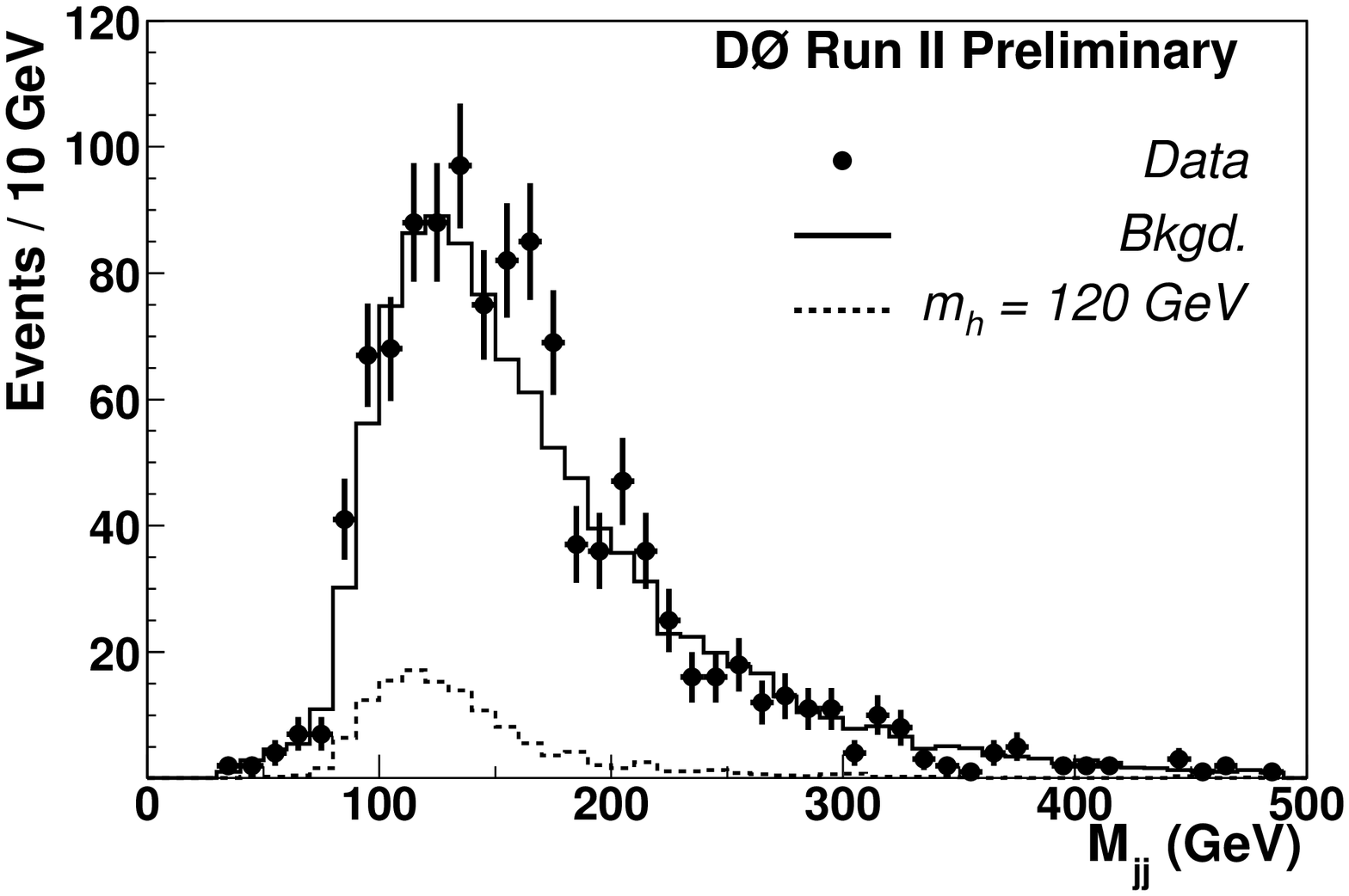}
\includegraphics*[width=0.48\textwidth]{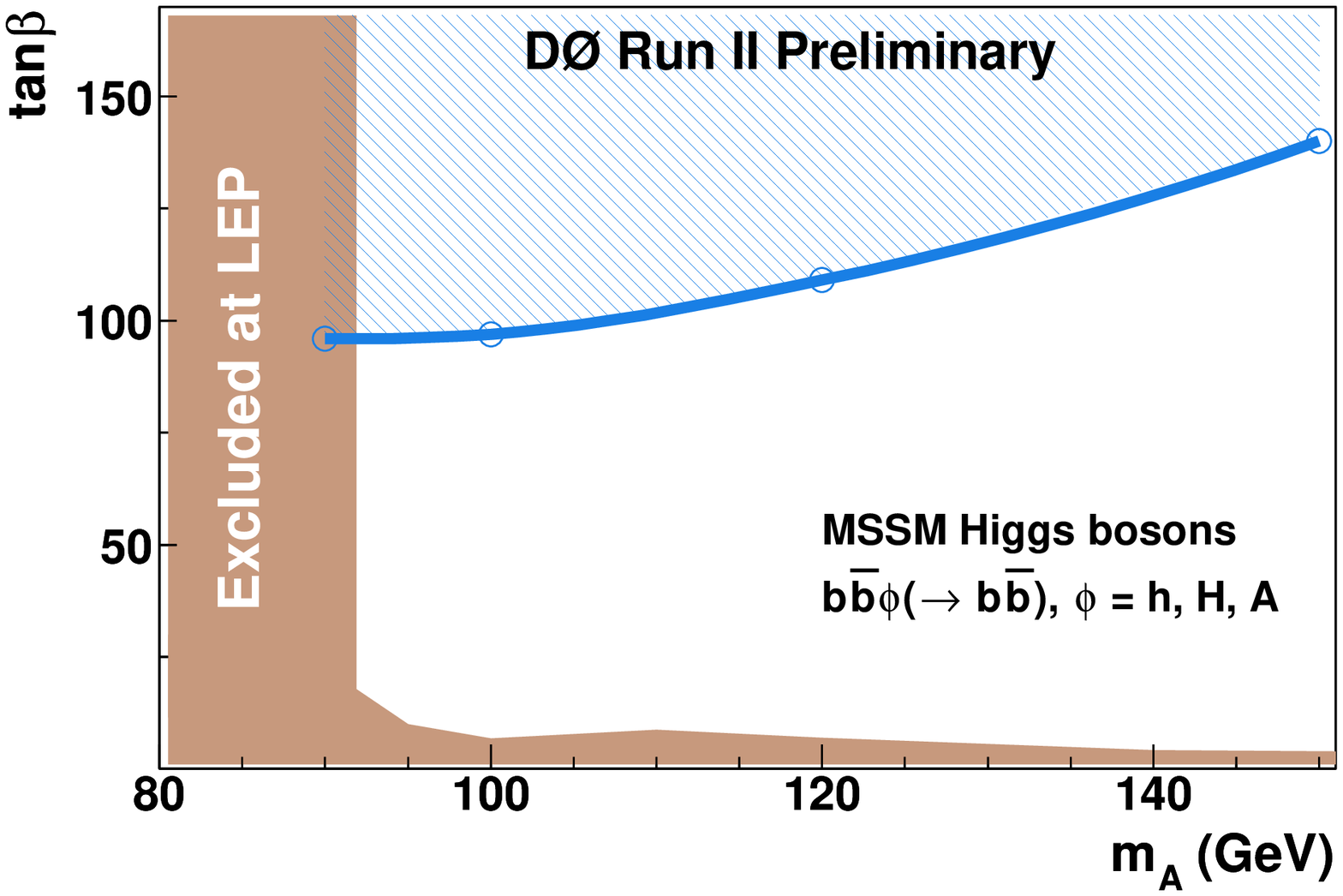}
\caption{%
Invariant mass spectrum of the two leading jets in the 3-jet sample
with three b-tags after final cuts in the D\O\ search for bh
production using 131~$pb^{-1}$ of Run~II data (left); Regions in (\tb,
m$_A$) excluded by this analysis at 95\% C.L. in comparison 
with LEP limits (right).
}
\label{f:tev-bbh}
\end{center}
\end{figure}

The background at this stage is dominated by multijet production with
b-quarks. Further discrimination is possible by 
searching for a peak in the invariant mass spectrum of the two leading
jets, which is shown in Fig.~\ref{f:tev-bbh} 
in comparison with the
expectation from background and a Higgs signal with m$_\hs$=120~GeV. 
The shape of the dijet mass spectrum in background is obtained from a
multijet sample with two b-tagged jets, which is expected to have
negligible contamination from signal, by weighting events using b-tag
fake rates measured in data as a function of jet \pt\ and~$\eta$.  
The background is then normalized by fitting this shape to the mass
spectrum outside the signal region. 

No evidence for production of neutral Higgs bosons~$h$, $H$, $A$ in 
association with b-jets has been observed, which allows to derive
limits on \tb\ as a function of m$_A$. In Fig.~\ref{f:tev-bbh}
the region excluded in the plane of (\tb, m$_A$) is shown 
in comparison with existing limits set by the LEP experiments.
The D\O\ Run~II limit is significantly worse than the 
limit published in 2001 by the CDF collaboration based on the analysis of 
91~\pbinv\ of Run~I data.\cite{tev-bbh-cdf}
Detailed comparisons of both results indicate that the apparent
loss of sensitivity observed by D\O\ 
can be traced back to the use of more recent cross-section calculations and PDF fits, 
which cause a significant reduction in signal acceptance and cross-section compared 
to the CDF Run~I analysis.\cite{tev-bbh-comparison}

With more luminosity and after combining results from both Tevatron
experiments, the reach in \tb\ at the 95\% C.L. will be extended to
about \tb=25 at m$_A$=120~GeV  
(for 5~\fbinv, within the {\it mhmax} scenario~\cite{mhmax}), but
deteriorates quickly with increasing m$_A$.  

\subsubsection{$\hs\to\tau\tau$}
In addition to the usually dominant decay mode $\hs\to\bb$, a light supersymmetric
Higgs boson can be searched for in its decay to $\tautau$.
This decay mode is of particular interest both for SUSY scenarios that favour
suppressed couplings of Higgs bosons to b-quarks as well as for the large
\tb\ region, where the channels $\hs b(b)\to \tau\tau b(b)$ and
$gg\to \hs\to\tau\tau$ provide a viable complement to the search for
$\hs b(b)\to bbb(b)$.

Both
Tevatron experiments have demonstrated the ability to reconstruct
hadronic tau decays in Run~II data by measuring the $Z\to\tau\tau$
cross-section.
The CDF collaboration has analyzed 200~\pbinv\ of Run~II data in search
for $gg\to \hs\to\tau\tau$ with one tau decaying leptonically to
electron or muon and the other tau decaying into
hadrons.\cite{tev-susyhiggs} 
The hadronic tau decay is reconstructed as one or more tracks pointing
to a narrow energy deposition in the calorimeter.  
Background from jets misreconstructed
as tau objects is further suppressed using cuts on track multiplicity,
mass and isolation of the tau candidate. 
The selection then 
requires one such tau candidate in addition to an isolated electron or muon.
After topological cuts using the transverse momenta of the lepton, the
tau candidate as well as the transverse missing energy, the sample is
dominated by irreducible background from $Z\to\tautau$ with a purity
of 90\%. Higgs events are selected with an  
efficiency of about 7\% (5\%) in the electron (muon) channel.

\begin{figure}[htb]
\begin{center}
\includegraphics*[width=0.46\textwidth]{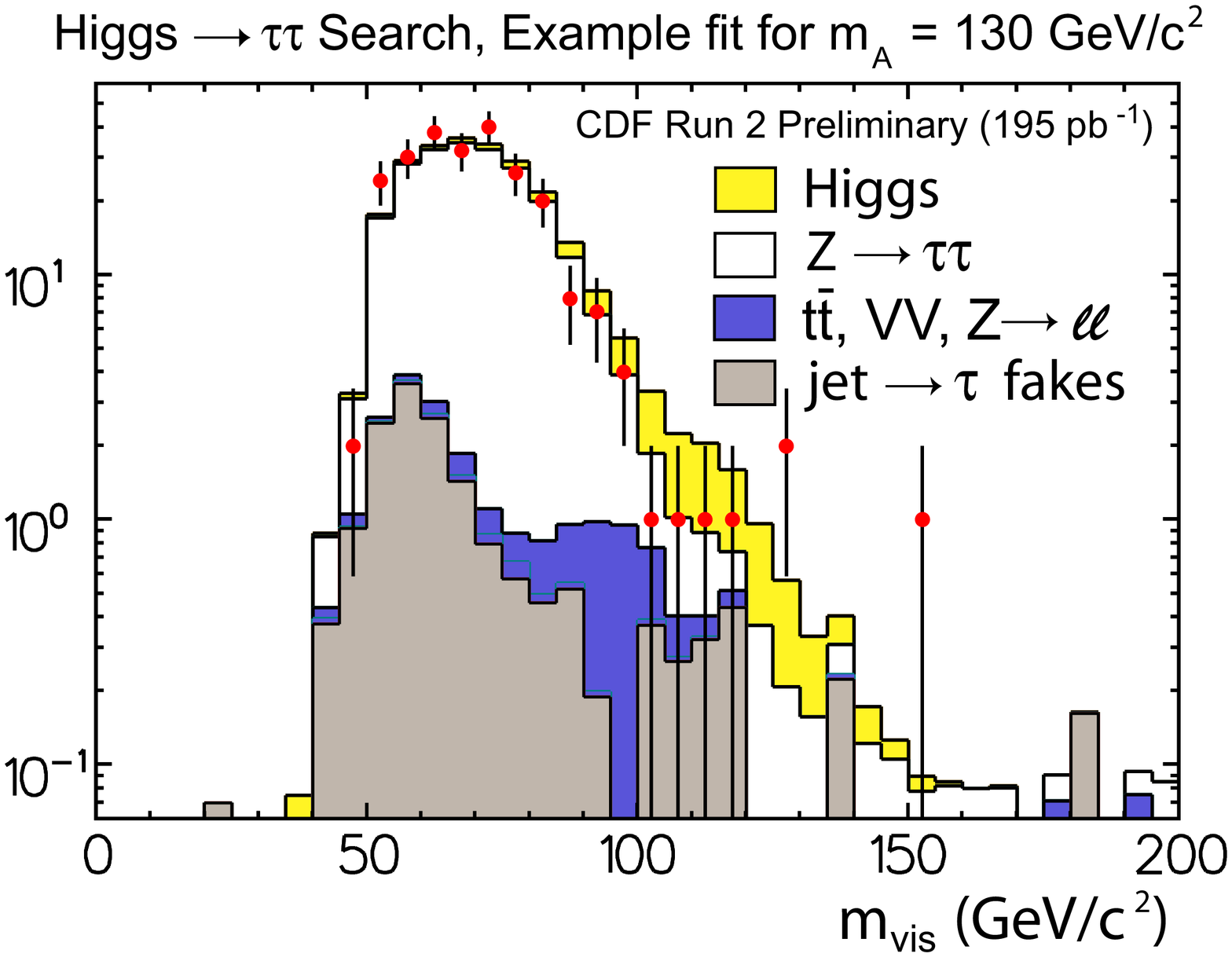}
\includegraphics*[width=0.50\textwidth]{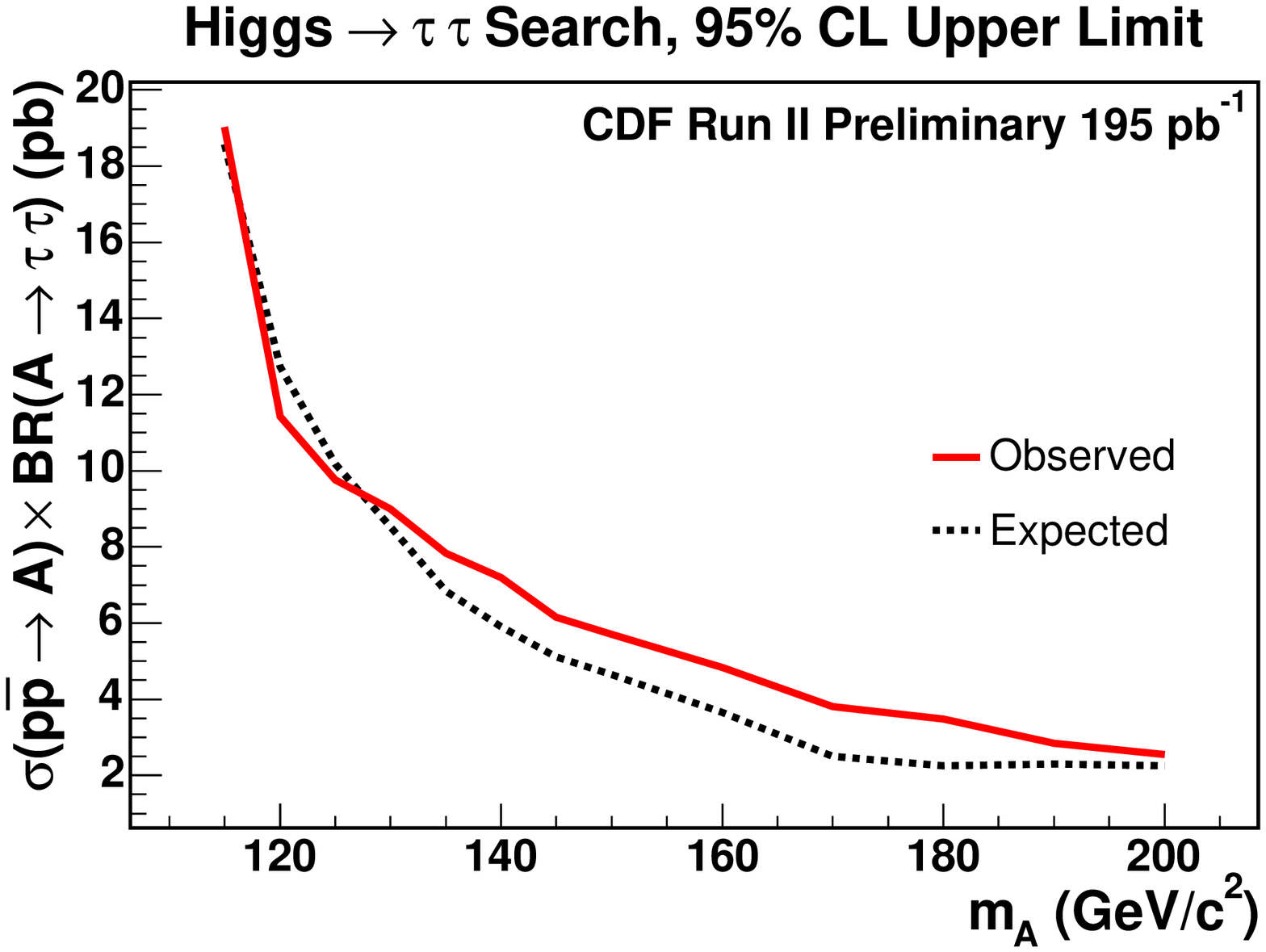}
\caption{%
Distribution of visible mass after all cuts of the CDF search for $H\to\tautau$
in 195~$pb^{-1}$ of Run~II data (left); Upper limit (at 95\% CL) on
the cross section $\sigma \cdot BR(h\to \tau\tau)$  in comparison with
the expected limit (right).
}
\label{f:tev-htautau}
\end{center}
\end{figure}

Separation of signal events from the $Z\to\tautau$ background is possible
by reconstructing an event mass~$m_{vis}$ based on the momenta of lepton and tau
candidate as well as the missing transverse energy.
Fig.~\ref{f:tev-htautau}a 
shows the distribution of $m_{vis}$ for data, backgrounds and a potential Higgs signal.
No evidence for an excess of events with respect to the Standard Model prediction
has been observed.
Using a binned likelihood fit of this distribution, a limit on the production
cross-section of $\hs\to \tau\tau$ has been extracted as displayed in
Fig.~\ref{f:tev-htautau}b as a function of the Higgs boson mass.

\subsubsection{Combined Reach}
\begin{figure}[htb]
\begin{center}
\includegraphics*[width=0.48\textwidth]{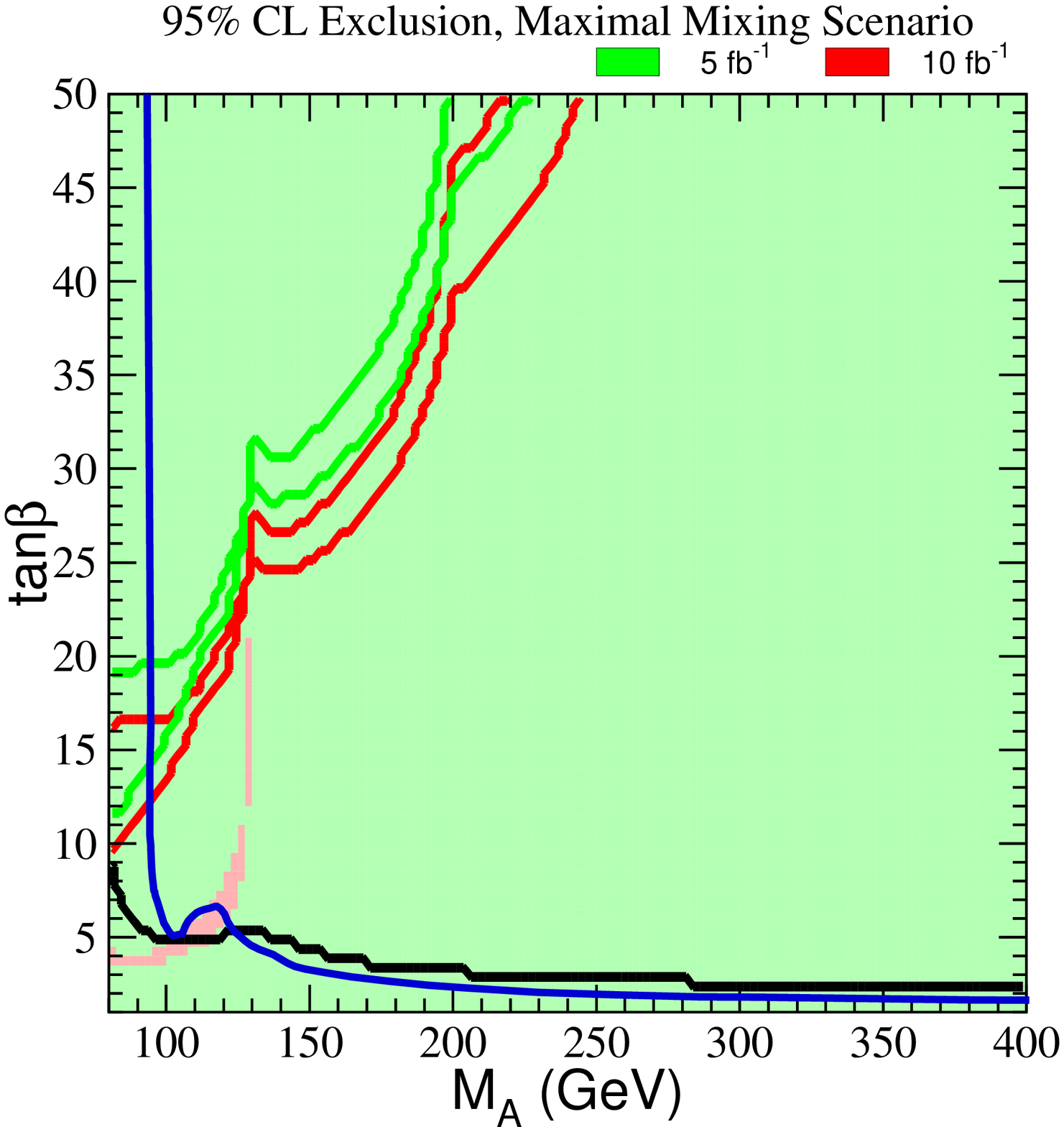}
\includegraphics*[width=0.48\textwidth]{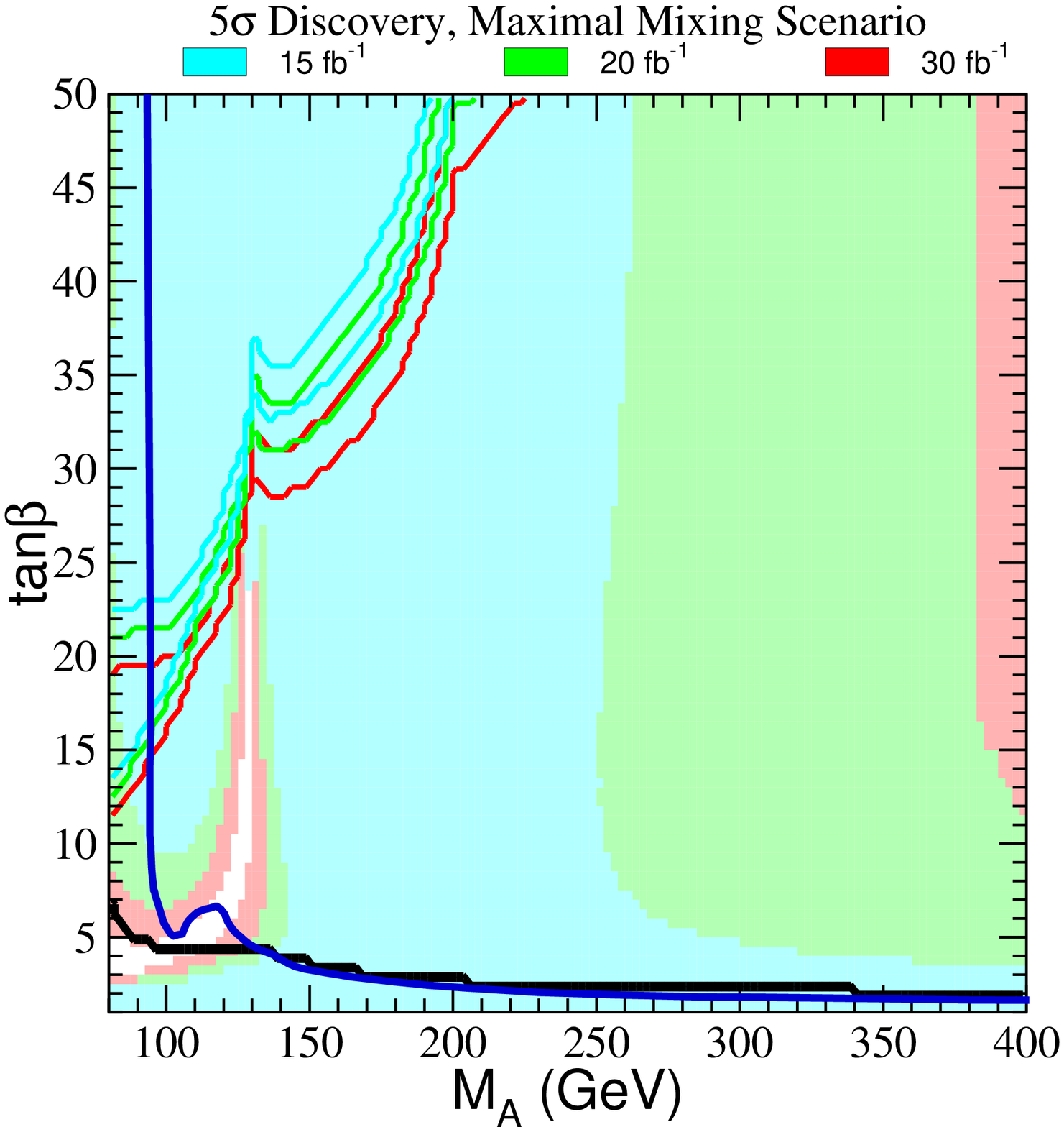}
\caption{%
Luminosity required for exclusion at 95\% C.L. (left) or 5$\sigma$
discovery (right) of a SUSY Higgs boson as a function of m$_A$ and
\tb\ within the mhmax scenario (taken from
Ref.~\protect\cite{tev-report} and modified to include most recent
LEP2 limit\protect\cite{lep-mssm}). Shaded regions indicate the
reach of $WH$/$ZH$ searches, the region above the diagonal lines are
accessible to searches for $hb(b)$, the dark line indicates the LEP2 limit.
}
\label{f:tev-susyreach}
\end{center}
\end{figure}

Combining dedicated searches for Higgs bosons at high \tb\ with 
searches for production of Higgs bosons in association with vector
bosons, sensitivity at 95\% C.L. to MSSM Higgs bosons within the 
{\it mhmax} scenario can be achieved 
independent of \tb\ with 5~\fbinv\ per experiment,
as shown in Fig.~\ref{f:tev-susyreach}.\cite{tev-report}
However, within this challenging scenario, a 5$\sigma$ discovery will 
not be possible at Tevatron Run~II for most of the (\tb, m$_A$)
plane. 

\subsection{Charged Higgs Bosons}
Models with an extended Higgs sector predict charged Higgs
bosons~$H^\pm$, or in the case of additional Higgs Tripletts
doubly-charged Higgs bosons~$H^{\pm\pm}$. 
Production of doubly charged Higgs bosons can provide particularly
striking signatures in LR inspired models, where BR($H^{\pm\pm}\to
\ell^\pm \ell^\pm$) is expected to be 100\%.  
CDF (D\O) have analyzed 240~\pbinv\ (107~\pbinv) of Run~II data to search for
$H^{\pm\pm}$ production in like-sign dilepton events.\cite{tev-susyhiggs} Requiring
two acoplanar, isolated electrons or muons, no excess of events has been
observed at high dilepton masses. For left-handed (right-handed)
$H^{\pm\pm}$, CDF set a lower mass 
limit of 135~GeV (112~GeV) for BR($H^{\pm\pm}\to \ell^\pm \ell^\pm$)=1.

\section{\label{susy}Searches for Supersymmetry}
Supersymmetry predicts a large number of new particles, most of which
could be light enough to be produced at the Tevatron. The CDF and D\O\
collaborations have searched their Run~II data for evidence of
squarks, gluinos, charginos and neutralinos.
For most analyses, minimal Supergravity (mSUGRA) is used as a
reference model for optimisation of the analysis and interpretation of
the result, even though the resulting  cross-section limits 
can be interpreted in a more model-independent way. 
Alternative models leading to different final state topologies have
been considered as well, including models with gauge-mediated SUSY 
breaking as well as R-parity violation.

The following sections summarize a selection of current Tevatron
results with relevance to Supersymmetry.

\subsection{Squarks and Gluinos}
\begin{figure}[htb]
\begin{center}
\includegraphics*[width=0.56\textwidth]{plots/N05F04.epsi} \hfill
\includegraphics*[width=0.39\textwidth]{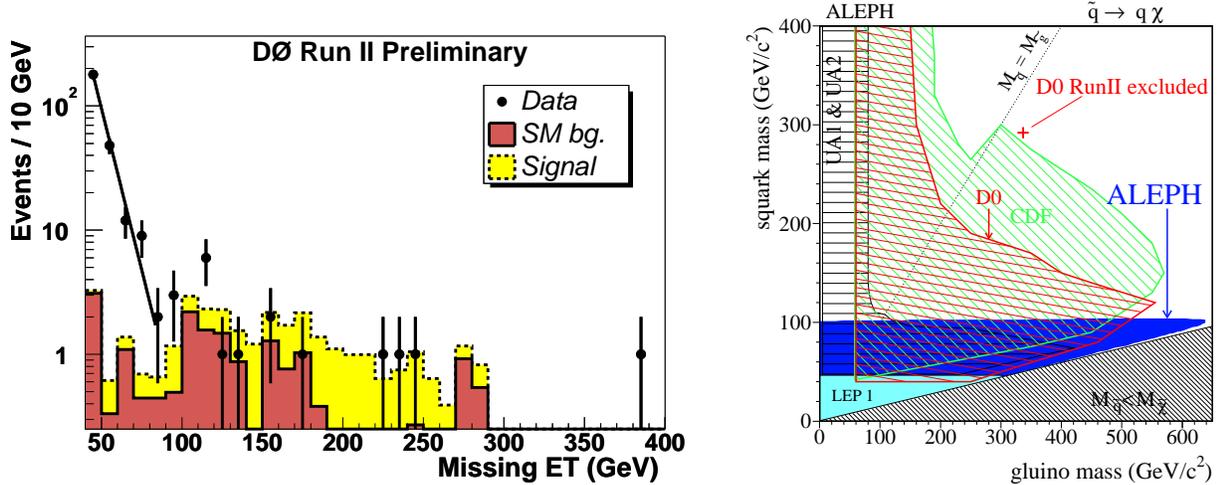}
\caption{%
Distribution of \met\ after all cuts except for the cut on \met\ itself (left), showing contributions from multijet backgrounds (line), other standard model processes (red) and signal (yellow); point in the squark/gluino mass plane corresponding to the limit set by the 
D\O\ analysis for the mSUGRA model line with $m_0$=25~GeV, \tb=3, $A_0$=0 and $\mu<0$ (right).
}
\label{f:squarks}
\end{center}
\end{figure}

Squarks and gluinos are produced through the strong interaction, resulting in 
relatively large signal cross-sections and therefore providing, if kinematically 
accessible, a promising signature for Supersymmetry at the Tevatron. 
The final state contains two or more jets along with missing transverse energy carried
away by the two lightest supersymmetric particles.
D\O\ have searched for pair production of squarks, each decaying into
a quark and the lighest neutralino.\cite{np-squarks}
This decay channel is expected to be dominant if the gluino is 
heavier than the squark.
85~\pbinv\ of Run~II data collected with a dedicated multijet trigger have been analyzed
using tight cuts on \met$>$175~GeV and $H_T$$>$275~GeV against the massive 
multijet background.
Fig.~\ref{f:squarks}a shows the \met\ distribution after all other cuts, with a tail
at high \met\ expected from standard model processes such as $Z$+jets with $Z\to\nu\bar{\nu}$.
After a veto of events containing isolated leptons to supress background from 
leptonic $W$ decays, an expectation of 2.7$\pm$1.0 Standard Model events remains.
Four events are observed in the D\O\ data, allowing to set upper limits on the squark 
production cross-section of about 2~pb.
Assuming that squarks are degenerate in mass for the first two generations, the
reach in mSUGRA parameter space has been evaluated for $m_0$=25~GeV, \tb=3, $A_0$=0 
and $\mu<0$ (see Fig.~\ref{f:squarks}). For this model line, squark masses below 292~GeV
can be excluded, corresponding to a slight improvement over existing 
limits (see Fig.~\ref{f:squarks}b).

\begin{figure}[htb]
\begin{center}
\includegraphics*[width=8cm]{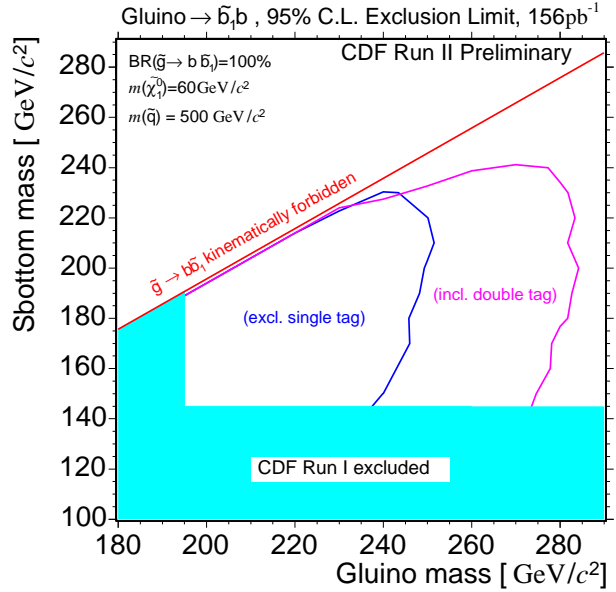}
\caption{%
Regions in the gluino/sbottom mass plane excluded by the CDF Run~II analysis requiring
one (blue line) or two (violet line) b-tagged jets.
}
\label{f:sbottom}
\end{center}
\end{figure}

For the third generation, mass unification is broken in many SUSY models due to
potentially large mixing effects. This can result in third generation squarks much 
lighter than the other squarks and the gluino.
The CDF collaboration has considered this scenario by searching 156~\pbinv\ of Run~II 
data for pair production of gluinos which subsequently decay to b-quarks 
plus sbottoms, resulting in final states with four b-quarks and missing transverse energy.\cite{np-sbottoms}
Signal events are isolated by requiring the presence of at least three jets with
\et$>$15~GeV, one or two of which need to be b-tagged using a secondary vertex algorithm.
After an additional cut on \met$>$80~GeV, 2.6$\pm$0.7 events are expected from Standard 
Model sources, while four events are observed in the data.
Assuming a gluino branching fraction to sbottoms of 100\%, regions in the gluino/sbottom mass plane 
can be excluded up to gluino masses of 280~GeV (see Fig.~\ref{f:sbottom}).

CDF also considered a scenario with light stop quarks, in this case assuming that
stops are long-lived and decay outside of the detector.\cite{np-stops}
Experimentally, pair production of light stable stop quarks can be detected as
a pair of heavily ionizing, slow moving charged particles. The events are triggered
on using standard muon triggers, and then selected in offline analysis using the 
time-of-flight detector. In this way backgrounds from light particles
are suppressed to an expectation of 2.9$\pm$0.7(stat)$\pm$3.1(syst) in 
53~\pbinv\ of Run~II data.
Seven events have been observed, allowing to exclude stop masses below 107~GeV (95~GeV) for isolated (non-isolated) stops.

\subsection{Charginos and Neutralinos}
\begin{figure}[htb]
\begin{center}
\includegraphics*[width=10cm]{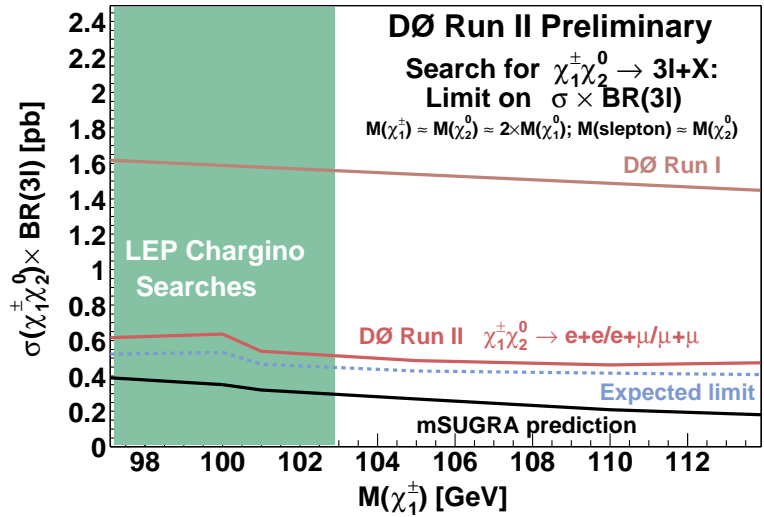}
\caption{%
The D\O\ Run~II limit (red line) and expected limit (dashed line) on cross-section 
times branching fraction into three leptons from the search for associated 
chargino/neutralino production. The shaded area is excluded by direct chargino searches
at LEP, the black line shows the expectation within mSUGRA for slepton masses equal 
to the second lightest neutralino mass.
}
\label{f:trileptons}
\end{center}
\end{figure}

Many SUSY models expect squarks and gluinos to be the heaviest supersymmetric particles, 
which might well put them out of the reach of the Tevatron Run~II experiments. In 
this case, a search for the associated production of charginos and 
neutralinos provides the most promising way for direct detection of supersymmetric
particles at the Tevatron. Due to its striking signature, the trilepton channel
$\cha^{\pm}\chiz_2\to 3l+\nu+\chiz_1\chiz_1$ is considered the most
powerful analysis channel, despite its low rate due to the small
cross-section and branching fraction.  
The D\O\ collaboration has searched for an excess of trilepton events in 175~\pbinv\
of Run~II data.\cite{np-trileptons}
Three different selections have been defined to cover topologies with
two electrons and a third isolated track, one electron plus one muon with a third 
isolated track, and two muons with the same charge.
To maintain the highest-possible efficiency, the third lepton is reconstructed (if at all)
as an isolated track, which is designed to be efficient for electrons,
muons and hadronic tau decays. 
A total of 0.9 events are expected from Standard Model backgrounds, dominated by
irreducible backgrounds from $WW$ and $WZ$ as well as $W\gamma$ with a converted photon.
Two events are observed in data, allowing to set an upper limit of about 0.6~pb 
on production cross-section times branching fraction into three leptons.
Fig.~\ref{f:trileptons} shows this limit as a function the chargino mass in comparison 
with the LEP limit from direct chargino searches.
The new D\O\ limit significantly improves on the Run~I limit, but for signal rates 
as expected within minimal SUGRA is not yet sensitive to chargino masses beyond
the LEP limit.

\subsection{$B_s\to\mu\mu$}
\begin{figure}[htb]
\begin{center}
\includegraphics*[width=7cm]{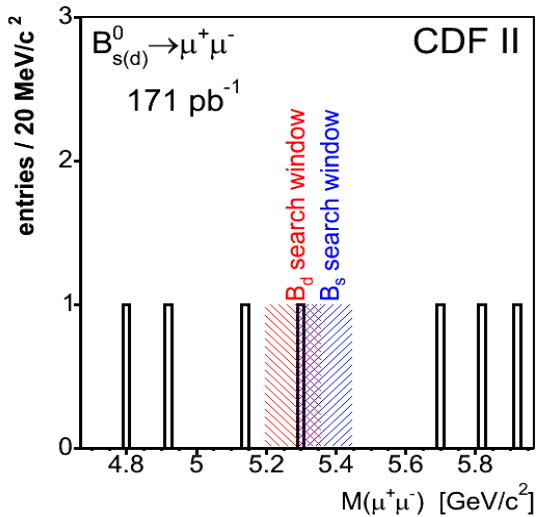}
\caption{%
Dimuon invariant mass distribution after all cuts of the CDF Search for $B_s\to\mu\mu$. 
One event is found in the $B_s$ search window, consistent with the background 
expectation extrapolated from the side bands.
}
\label{f:bsmumu}
\end{center}
\end{figure}

While not a direct search for supersymmetric particles, the measurement of the 
branching fraction of $B_s\to\mu\mu$ is a sensitive probe of
Supersymmetry. This rare flavour-changing neutral current 
decay is heavily suppressed within the Standard Model, where a branching fraction 
of only 3.8$\times 10^{-9}$ is expected.
However, within supersymmetry this decay can be significantly enhanced by loop
corrections. For instance, within SUGRA the branching fraction is proportional 
to (\tb)$^6$, leading to an enhancement of up to three
orders of magnitude.\cite{Bs-sugra}
At the Tevatron, $B_s$ mesons are produced with a very large rate, and a possible decay
into two muons can be identified with high efficiency. Both collaborations have reported
on analyses
searching about 170~\pbinv\ of Run~II data for evidence of this rare decay. The selection
suppresses dimuon background from prompt production and semi-leptonic b-decays
by requiring two isolated muons originating from a vertex that is significantly 
displaced from the primary interaction point.
Both CDF and D\O\ estimate the sensitivity of their analysis to branching fractions 
of about 9$\times 10^{-7}$ at 95\% C.L.\cite{bsmumu}
While D\O\ have not yet quoted a result, 
CDF observe no significant excess of events in the search
window around the $B_s$ mass (see Fig.~\ref{f:bsmumu}) and have set a limit of 
BR($B_s\to\mu\mu$)$<$7.5$\times 10^{-7}$ at 95\% C.L.

\subsection{Gauge-Mediated SUSY Breaking}
\begin{figure}[htb]
\begin{center}
\includegraphics*[width=0.50\textwidth]{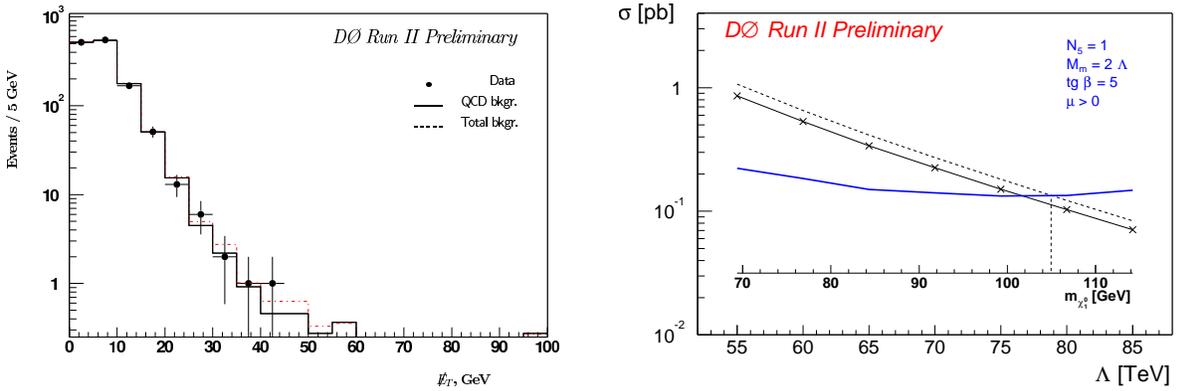}
\includegraphics*[width=0.46\textwidth]{plots/N04F03.epsi}
\caption{%
The distribution of missing transverse energy (left) after all cuts of the D\O\ search for 
GMSB SUSY, compared to the expectation within the Standard Model (dashed line) and the prediction for multijet events (solid line); Upper cross-section limit on GMSB SUSY with neutralino NLSP in comparison with the NLO signal cross-section (right).
}
\label{f:gmsb}
\end{center}
\end{figure}

The mass hierarchies expected in models with gauge-mediated SUSY
breaking (GMSB) can 
significantly alter the signatures of supersymmetry at colliders. 
While the Gravitino is expected to be the lightest supersymmetric particle
in these models, the NLSP can be either a neutralino or a slepton.
Neutralino NLSPs decay to a photon and a gravitino, the latter escaping detection in a 
collider detector. Production of charginos and neutralinos therefore leads to final
states containing at least two photons and missing transverse energy.
Both Tevatron collaborations have searched for an excess of such events in about
200~\pbinv\ of Run~II data.\cite{np-gmsb}
After requiring missing transverse energy larger than
40~GeV (45~GeV), D\O\ (CDF) observe one (zero) events compared to an expected 
background of 2.5$\pm$0.5 (0.6$\pm$0.5) events. 
Limits on the production cross-section of charginos and neutralinos have been set
as a function of chargino mass (see Fig.~\ref{f:gmsb}).
Within a particular GMSB scenario (one  messenger field, M=2$\Lambda$, \tb=5, $\mu>$0), these limits can be translated into lower limits on chargino (neutralino) masses of
192~GeV (105~GeV) and 168~GeV (93~GeV) for D\O\ and CDF, respectively.

\section{\label{other}Other Searches for New Physics}
A large variety of other searches have been performed at the Tevatron.
While not directly aimed at the discovery of supersymmetric particles,
most of the topologies covered by these searches are of relevance for
SUSY models as well. This section gives a brief overview of these results.

\begin{figure}[htb]
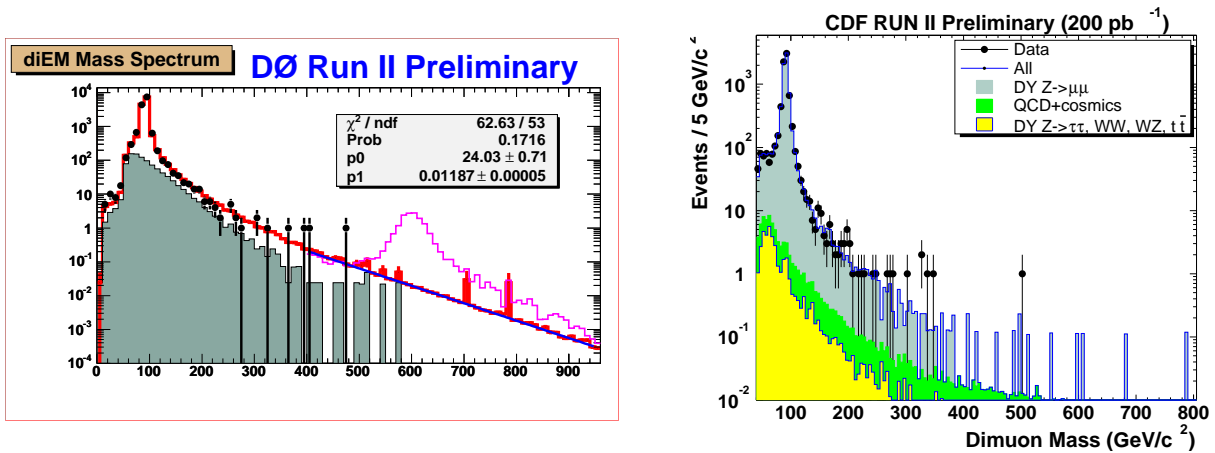

\begin{center}
\begin{minipage}{0.51\textwidth}
\includegraphics*[width=\textwidth]{plots/N03F01.epsi}
\end{minipage}  \hfill
\begin{minipage}{0.43\textwidth}
\includegraphics*[width=\textwidth]{plots/data_winter04_all_log_2.epsi}
\end{minipage}
\caption{%
Distribution of invariant dielectron (left) and dimuon (right) mass in
the high-pt dilepton searches by D\O\ and CDF, respectively. The left
plot shows the MC expectation for a $Z'$ signal with m$_{Z'}$=600~GeV on
top of the Standard Model backgrounds.
}
\label{f:zprime}
\end{center}
\end{figure}

Final states with two high-pt leptons or photons are predicted in a
number of extensions of the Standard Model. Signals include new
neutral gauge bosons~$Z'$ as well as the  production or exchange of
gravitons in models with extra dimensions.
Both CDF and D\O\ have observed no significant excess in their high-pt
dilepton and diphoton data, as demonstrated by the good agreement
between data and background predictions in the invariant dilepton
mass distributions shown in Fig.~\ref{f:zprime}.
For given models, these results can be translated into limits on $Z'$
masses or the fundamental Planck scale as summarized in
Table~\ref{limits}.\cite{np-dileptons} 
\begin{table}[hbt]
\begin{center}
\caption{
Selection of most stringent CDF and D\O\ limits on dilepton
resonances, leptoquarks, excited leptons and large extra
dimensions. The results are quoted for $(1)$ a sequential $Z'$ with
Standard Model couplings, $(2)$ scalar leptoquarks with 100\%
branching fraction, $(3)$ a contact interaction scale equal to the
mass of the excited electron, $(4)$ the GRW convention, $(5)$ four
extra dimensions. 
}
\begin{tabular}{|lrr|c|r|}
\hline
\multicolumn{3}{|c|}{Analysis \rule[-1.5ex]{0ex}{4.5ex}} 
 & \multicolumn{1}{c|}{\ \ Channel\ \ }  &  \multicolumn{1}{c|}{Limit}\\
\hline
High-mass Dilepton & D\O, & 200~\pbinv\ & $Z'\to ee$ & m$_{Z'}>$780~GeV$^{(1)}$\\
 & CDF, & 200~\pbinv\ & $Z'\to\mu\mu$ & m$_{Z'}>$735~GeV$^{(1)}$\\
 & CDF, & 195~\pbinv\ & $Z'\to\tau\tau$ & m$_{Z'}>$394~GeV$^{(1)}$\\
\hline
Leptoquarks & D\O, & 175~\pbinv\ & $LQ\to eq$ & m$_{LQ}>$240~GeV$^{(2)}$\\
 & CDF, & 198~\pbinv\ & $LQ\to$$\mu q$ & m$_{LQ}>$241~GeV$^{(2)}$\\
 & CDF, & 191~\pbinv\ & $LQ\to$$\nu q$ & m$_{LQ}>$117~GeV$^{(2)}$\\
\hline
Excited Electrons & CDF, & 200~\pbinv\ & $e^*\to e\gamma$ & m$_{e^*}>$889~GeV$^{(3)}$\\
\hline
Large Extra Dimensions & D\O, & 200~\pbinv\ & $\gamma\gamma$, $ee$ & M$_S$$>$1.43~TeV$^{(4)}$\\
& D\O, & 85~\pbinv\ & jet+$G_{KK}$ & M$_D$$>$685~GeV$^{(5)}$\\
\hline
\end{tabular}
\label{limits}
\end{center}
\end{table}

Similar searches exist for other new
high-mass particles such as leptoquarks~\cite{np-leptoquarks}, 4th generation
quarks~\cite{np-tprime} and excited leptons~\cite{np-eleptons}.
Given the high mass scale of the signal, fairly stringent cuts on
transverse energy can be applied to select final states with high-pt
jets, leptons or large missing transverse energy. 
No evidence for any excess has been reported. Table~\ref{limits}
summarizes the mass limits that have been derived by CDF and D\O.

\section{Acknowledgements}
I would like to thank my colleagues at CDF and D\O\ for providing the material for this presentation as well as the organizers of the SUSY~2004 conference for this very well-organized event.

\bibliographystyle{plain}

\end{document}